%% file: main.tex
\documentclass[conference,compsoc]{IEEEtran}
\ifCLASSOPTIONcompsoc
  \usepackage[nocompress]{cite}
\else
  \usepackage{cite}
\fi
\ifCLASSOPTIONcompsoc
\usepackage[caption=false,font=footnotesize,labelfont=sf,textfont=sf]{subfig}
\else
 \usepackage[caption=false,font=footnotesize]{subfig}
\fi
\usepackage{nopageno}

\usepackage[normalem]{ulem}

\usepackage{multicol}
\usepackage{graphicx}
\usepackage{cite}
\usepackage[breaklinks, bookmarks=false]{hyperref}
\hypersetup{
    colorlinks,
    citecolor=black,
    filecolor=black,
    linkcolor=black,
    urlcolor=black
}
\usepackage{tikz}
\usepackage{amsmath}
\newcommand{\paragraphb}[1]{\vspace{0.07in}\noindent{\bf #1} }

\usepackage{soul}
\usepackage{enumitem}

\newenvironment{myquote}%
  {\list{}{\leftmargin=0.1in\rightmargin=0.1in}
  \item[]}%
  {\endlist}
\AtBeginEnvironment{myquote}{\itshape}

\begin{document}
\title{Understanding Parents' Perceptions and Practices\\Toward Children's Security and Privacy in Virtual Reality}

\author{
\IEEEauthorblockN{Jiaxun Cao}
\IEEEauthorblockA{Duke University \\
jessie.cao@duke.edu}
\and
\IEEEauthorblockN{Abhinaya S.B.}
\IEEEauthorblockA{North Carolina State University \\
asrivid@ncsu.edu}
\and
\IEEEauthorblockN{Anupam Das}
\IEEEauthorblockA{North Carolina State University \\
anupam.das@ncsu.edu}
\and
\IEEEauthorblockN{Pardis Emami-Naeini}
\IEEEauthorblockA{Duke University \\
pardis@cs.duke.edu}
}

\maketitle
\thispagestyle{plain}
\pagestyle{plain}

\begin{abstract}
\input{sections/00-abstract}
\end{abstract}

\IEEEpeerreviewmaketitle

\input{sections/01-introduction}
\input{sections/02-relatedwork}

\input{sections/03-methods}
\input{sections/04-results}

\input{sections/05-discussion}

\section{Conclusion} \label{sec:conclusion}
\input{sections/06-conclusion}

\section*{Acknowledgments}
We thank Sparsha Perupalli, anonymous reviewers, and the shepherd for their constructive feedback. We also thank Iffat Anjum and Nazanin Sabri for helping with our pilot interviews. We further thank our participants for their insightful input. This material is based upon work supported in parts by a Meta Research gift award. Any opinions, findings, conclusions, or recommendations expressed in this material are those of the authors and do not necessarily reflect the views of Meta Inc.

\bibliographystyle{IEEEtranS}
\bibliography{references}

\input{sections/appendix.tex} 

\end{document}

%% file: sections/00-abstract.tex
Recent years have seen a sharp increase in the number of underage users in virtual reality (VR), where security and privacy (S\&P) risks such as data surveillance and self-disclosure in social interaction have been increasingly prominent. Prior work shows children largely rely on parents to mitigate S\&P risks in their technology use. Therefore, understanding parents' S\&P knowledge, perceptions, and practices is critical for identifying the gaps for parents, technology designers, and policymakers to enhance children's S\&P. While such empirical knowledge is substantial in other consumer technologies, it remains largely unknown in the context of VR. To address the gap, we conducted in-depth semi-structured interviews with 20 parents of children under the age of 18 who use VR at home. Our findings highlight parents generally lack S\&P awareness due to the perception that VR is still in its infancy. To protect their children's interactions with VR, parents currently primarily rely on active strategies such as verbal education about S\&P. Passive strategies such as using parental controls in VR are not commonly used among our interviewees, mainly due to their perceived technical constraints. Parents also highlight that a multi-stakeholder ecosystem must be established towards more S\&P support for children in VR. Based on the findings, we propose actionable S\&P recommendations for critical stakeholders, including parents, educators, VR companies, and governments.

%% file: sections/01-introduction.tex
\section{Introduction}
\label{sec:intro}
The growing popularity of consumer-facing virtual reality (VR) in recent years has attracted a significant number of users~\cite{bagheri2016virtual, epp2021empirical}, including children under the age of 18~\cite{maloney2020complicated, deldari2023investigation, maloney2020virtual}. While immersive and realistic experiences provide great user engagement, more risks have been emerging in VR, including safety, well-being, security, and privacy (S\&P) issues~\cite{adams2018ethics}. Specifically, VR devices collect a variety of sensitive data~\cite{dick2021balancing}, including eye- and motion-tracking data that are indispensable for immersion yet pose risks of data surveillance~\cite{dick2021balancing, caserman2019real, clay2019eye}. Worse yet, S\&P experts have been warning people of the scarce, unspecific, and under-enforced privacy policies in VR~\cite{Kelly_2022, Reed_2022, Klein_2022, dick2021balancing}.

Furthermore, users including children have flooded social VR applications (e.g., Horizon Worlds, Rec Room, and VRChat)~\cite{maloney2020virtual, maloney2020complicated, deldari2023investigation}, raising S\&P concerns. In internet-enabled social VR applications, besides data surveillance, users also face S\&P risks in social interaction, including self-disclosure~\cite{maloney2020anonymity}. Prior work has suggested that social VR users easily disclose a wide range of personal information to strangers, including emotions, life experiences, voices, and even real-life appearances~\cite{maloney2020anonymity}.

In particular, these S\&P risks in VR can be heightened for children, especially considering the largely age-mixed environment in VR~\cite{deldari2023investigation, maloney2020complicated, maloney2020virtual}. Prior work has investigated common risks that children encounter in social VR, including virtual harassment, cyberbullying, and inappropriate content from both peers and adults~\cite{deldari2023investigation, maloney2020complicated, maloney2020virtual}. However, one important limitation of the existing literature is that while general safety risks have been widely documented, S\&P-specific risks in VR have been largely overlooked.

In mitigating S\&P risks in technologies, children heavily rely on parents' S\&P knowledge and decision-making~\cite{sorensen2016protecting, livingstone2019children, kumar2017no}. Researchers have suggested that children trust parents' S\&P rules and tend to seek help from parents when risks emerge~\cite{kumar2017no}. Moreover, parents' S\&P perceptions might influence how children conceptualize and perceive S\&P~\cite{shin2012tweens, chai2009internet, boyd2011social}. In addition, parents also adopt a variety of strategies when mediating children's technology use, including passive strategies (e.g., parental controls) and active strategies (e.g., verbal education)~\cite{kumar2017no, zhang2016nosy, wisniewski2015preventative, wisniewski2017parental}. When adopting different strategies, parents might also face different challenges in mitigating risks, such as low familiarity with technology~\cite{cranor2014parents, wisniewski2014adolescent}. Therefore, understanding parents' S\&P perceptions and practices is crucial for researchers and other important stakeholders to support parents' risk mitigation strategies. While such knowledge regarding other technologies (e.g., smartphones~\cite{hiniker2016not, akter2023takes}, Internet-enabled applications~\cite{kumar2017no, akter2022parental}) has been fully investigated, it remains unknown in the context of VR. Hence, this work aims to answer three research questions (RQs):

\begin{itemize}[leftmargin=*]\itemsep=0pt
    \item \textbf{RQ1}: What are parents’ perceptions of children’s S\&P in VR?
    \item \textbf{RQ2}: What are parents’ risk mitigation strategies for their children’s S\&P in VR?
    \item \textbf{RQ3}: What are parents' expectations toward perceived critical stakeholders and future S\&P-enhancing features in VR?
\end{itemize}

To address the RQs, we recruited 20 parents whose children use VR applications at home. We conducted semi-structured interviews to surface their perceptions of S\&P risks for children in VR. We then asked parents how they addressed their concerns, if at all. Last, we asked them to propose critical stakeholders and desired features in protecting children's S\&P in VR.

Our qualitative analysis reveals three main points. (1) First, although parents typically do not have immediate S\&P concerns for their own children, they do express concerns for underage VR users in general and anticipate more pressing risks for their own children in the future. (2) Second, parents adopt a limited range of strategies when mediating children's VR usage. Specifically, parents rely on active strategies such as verbal education more than passive ones, including parental controls. While acknowledging the potential effectiveness of passive strategies, parents face significant technical constraints in practice. (3) Finally, parents agree on four critical stakeholders -- parents, educators, VR companies, and governments. They also propose a series of potential features to enhance existing S\&P controls in VR, with a few reservations about using them.

Based on parents' rich insights, we provide actionable recommendations for the critical stakeholders. In brief, we urge parents, educators, and governments to enhance their nuanced understanding of VR technologies, as their current S\&P practices are overgeneralized across all digital devices. We also encourage parents and educators to set physical references when educating children about S\&P risks in VR. In addition, we propose that VR companies should involve engagement, granularity, and modality-specific considerations in their designs of S\&P controls. Finally, we call for VR companies to enhance their user guidance on S\&P risks in VR, along with regulators' enhanced supervision over VR companies' S\&P practices.

%% file: sections/02-relatedwork.tex
\section{Related Work} 
\label{sec:relatedwork}


\paragraphb{S\&P risks in VR.} \label{sec:rw1}
VR technologies collect a vast amount of sensitive data that has been categorized by Dick~\cite{dick2021balancing} into four types -- (1) observable data that generate users’ virtual presence, e.g., avatars and real-time in-world interactions~\cite{o2016convergence, heller2020reimagining}; (2) observed data that enhance immersive experiences, e.g., motion/eye-tracking~\cite{caserman2019real, clay2019eye, heller2020reimagining, bar2019eyes, kroger2020does}; (3) computed data that improve user services, e.g., user profiles for advertisements; and (4) associated data that allow Internet-enabled functions, e.g., IP addresses, usernames, and friend lists~\cite{acosta2020ip}. Unlike many other consumer technologies, VR collects these sensitive data that are indispensable for its core utility -- immersive experiences~\cite{initiative2020xrsi}. Specifically, users have to trade their motion-tracking data for the full-body tracking experience; the data collected via external-facing cameras for detecting obstacles around; and voice data for social experiences~\cite{dick2021balancing}; just to name a few. Due to the unique immersive nature of VR technologies, researchers have been calling for policymakers to reform existing privacy laws and policies such as COPPA~\cite{pearlman2021securing} and HIPAA~\cite{gallardo2023speculative} that would unnecessarily limit users’ experiences of VR technologies~\cite{dick2021balancing}. In addition, a 2018 study has revealed that VR developers considered VR privacy policies problematic for providing developers with limited standards and guidelines regarding S\&P~\cite{adams2018ethics}.

In the context of social interaction, more S\&P risks have emerged as VR evolves to become a popular form of social technology~\cite{maloney2020anonymity, zheng2023understanding, sykownik2022something}. Mainstream social VR platforms, including VRChat, Horizon Worlds, Rec Room, and more, allow users to socialize with others through head-mounted displays (HMDs)~\cite{mcveigh2019shaping, mcveigh2018s}. In this vein, a volume of social VR work has identified numerous privacy and safety risks. Maloney et al.~\cite{maloney2020anonymity} have found that social VR users tend to feel comfortable and safe to share emotions and life experiences due to the anonymity afforded by social VR. In addition, users’ physical traits, such as their voices and real-life appearances, can also be disclosed since (1) voice chats are ubiquitous in social VR and (2) some users prefer to represent themselves with a virtual avatar that looks similar to their appearance in real life. More recently, with the increasing popularity of social VR streaming, users’ privacy has been violated by streamers who broadcast in public VR spaces~\cite{zheng2023understanding}.

In systematizing VR threats from the existing literature, a recent SoK paper~\cite{garrido2023sok} has characterized adversary types corresponding to four data-processing entities in VR -- hardware, client, server, and user adversaries. Hardware adversaries can access users' raw sensor data and manipulate information provided to VR applications and users. Client adversaries characterize how developers of VR applications and content creators may convey ``misinformative, manipulative, and deceptive content'' to users. Server adversaries can enable multi-user functionality and process networked data presented to users. Finally, user adversaries represent other users in the same VR application. After characterizing this threat model, Garrido et al.~\cite{garrido2023sok} have pointed out that user adversarial attacks easily occur, whereas hardware adversaries are still not mature. Importantly, due to the lack of privacy policy standards and enforcement~\cite{trimananda2022ovrseen}, server adversaries are also easily executable, inferring many sensitive data such as user demographics in a short time.


\paragraphb{VR risks for children.} Due to the age-mixed environment~\cite{deldari2023investigation, maloney2020complicated, maloney2021stay}, children who are considered part of the marginalized community in VR often face more tailored risks, such as predators, harassment, cyberbullying, inappropriate adult content~\cite{maloney2020virtual, blackwell2019harassment, shriram2017all, maloney2020talking, sabri2023challenges}, etc. However, compared to such safety threats that have been widely studied, children’s S\&P in social VR have been much more marginalized in scholarly work. Previous studies have collectively pointed out that bystanders and parents have concerns for children’s privacy as they interact with strangers in VR~\cite{deldari2023investigation, maloney2020complicated}. Notably, one study has suggested VR privacy guidelines be made more straightforward to align with children’s reading and literacy levels~\cite{maloney2020virtual}. To further provide actionable S\&P-enhancing recommendations, more research centered on children’s S\&P is needed. Since children heavily rely on parents’ decision-making of their online S\&P~\cite{sorensen2016protecting, youn2008parental, livingstone2019children, shin2016adolescents, kumar2017no}, a critical gap exists in terms of parents’ attitudes and knowledge around children’s S\&P in VR. 

\paragraphb{Parents' risk mitigation strategies toward technologies at home.}
Nissenbaum defines privacy as “neither a right to secrecy nor a right to control, but a right to appropriate flow of personal information.”~\cite{nissenbaum2020privacy}. This definition is especially applicable to online social contexts where privacy risks arise from self-disclosure~\cite{kumar2018co, nissenbaum2020privacy, joinson2007self, kramer2020mastering, hallam2017online, krasnova2009won}. When self-disclosure occurs online, privacy is about the appropriate flow of information~\cite{nolan2011stranger}. With children’s significantly increased online activities~\cite{subrahmanyam2000impact, livingstone2017children, livingstone2014annual}, more unexpected information flows have raised new privacy concerns~\cite{kumar2017no, palen2003unpacking}. For instance, information disclosed via email can be retained longer than face-to-face communication~\cite{kumar2017no}. It is critical to protect children’s privacy online since (1) privacy is an important space for children to practice their decision-making, create boundaries, and foster independence~\cite{kumar2018co, nolan2011stranger}; (2) privacy can benefit children’s experience in identity play, relationship building, and peer communication~\cite{livingstone2006children, kumar2017no}.

Security is another important aspect of children’s interactions with technology. Prior work suggests three goals of information security~\cite{kumar2017no, smith2011information}: (1) maintaining the integrity of personal information stored and transmitted through networks, (2) authenticating users with access to personal information, and (3) preserving the confidentiality of personal information. This work uses the above-mentioned definitions of S\&P to unpack parents’ risk perceptions and mitigation strategies.

Given the importance of S\&P for children’s technology use, parents have the responsibility in protecting children’s digital S\&P at home~\cite{sun2021they, shmueli2010privacy}. Kumar et al.~\cite{kumar2017no} have found that children aged 9 years old or younger are most likely to seek help from parents when confronted with unknown risks online and largely rely on parents’ explicit rules to deal with their digital S\&P risks. Other research in this vein has suggested that children of different ages trust their parents’ S\&P protections and guidance such as on data sharing~\cite{livingstone2014developing, yip2019laughing, sun2021they}. Additionally, another line of research has suggested that parents’ S\&P attitudes and practices may shape that of children's~\cite{sun2021they, feng2014teens, wisniewski2015preventative, bowyer2018understanding, shin2012tweens, chai2009internet, boyd2011social}. For example, Shin et al.~\cite{shin2012tweens} have reported that parents who self-identify as more S\&P-focused also tend to better direct children’s S\&P behavior online.

Parents adopt a variety of strategies to mitigate the risks of children’s technology use~\cite{kumar2017no}, including talking to children, viewing children’s accounts, using parental controls, being around their children during their technology use, setting rules and boundaries, etc.~\cite{duerager2012can, madden2012parents, zhang2016nosy, wisniewski2015preventative}. Having examined different parental strategies, Wisniewski et al.~\cite{wisniewski2017parental} have summarized three types of strategies in the Teen Online Safety Strategies (TOSS) framework: (1) passive monitoring, (2) rules and restrictions in usage, and (3) active mediation such as through having conversations with children. However, when mitigating the risks, parents face a number of challenges, such as underestimating the potential risks faced by children~\cite{wisniewski2017parents, blackwell2016managing, cranor2014parents}, undermining parent-child trust~\cite{cranor2014parents, hawk2009mind, livingstone2013regulating, smetana2008s}, and an incomplete understanding of technologies compared to their children~\cite{cranor2014parents, teitell2013digital, wisniewski2014adolescent}. Therefore, it is crucial to investigate parents’ S\&P practices around children’s technology use at home and identify the tradeoffs in their decision-making to help parents and technology designers work together towards safer and more inclusive technologies for children.

%% file: sections/03-methods.tex
\section{Methods}
\label{sec:method}
\paragraphb{Recruitment.} In July 2023, we advertised our study on VR-related Discord servers, sub-Reddits, LinkedIn, and Facebook groups. Prior to the advertisement, we obtained permission from all the channels' moderators, admins, or owners. In addition, one of the researchers reached out to a local school via personal connections. After we obtained the approval from the school, we advertised through the school's official email communication. When advertising our study, we did not specify our S\&P focus because we did not want to prompt our participants prior to the interview. We were interested in what concerns would first occur to parents in a natural response without preparation beforehand. Similar to Kumar et al.'s~\cite{kumar2017no} approach, we framed it as an investigation into parents' perceptions and practices of children's VR usage. In our advertisement posts and emails, we attached a link to the pre-screening survey and asked all interested candidates to fill out the survey.

\paragraphb{Eligibility and pre-screening.} To qualify for our study, participants must be (1) adult residents of the US, (2) a parent or legal guardian of a child, and (3) in possession of a VR headset in their household. The remainder of the pre-screening survey consisted of questions about parents’ and their children’s VR usage, including purposes of usage, VR headsets, and apps that they use. We then asked them basic demographic questions, including parents' and their children's age ranges, genders, ethnicities, and parents' highest degrees. We closed the pre-screening survey with email collection for further contact.

Based on the responses, we sent out interview invitations to qualified candidates on a rolling basis until we finished 20 valid interviews. In particular, we excluded candidates who or whose children had never used VR, e.g., only the candidate's partner had used it. In doing so, we attempted to interview participants with more experience in VR and mediation of children's VR usage.

\paragraphb{Interview procedure.} We emailed each participant one day before their scheduled interview to confirm their consent to participation. To verify the authenticity of participants, we began our interview by asking participants to show their VR headsets on camera. As a result, there were nine no-shows or withdrawals from participation. After each no-show or withdrawal, we emailed them to confirm if they would like to reschedule. Meanwhile, we sent out new invitations to other qualified candidates from our survey.

From July to September 2023, we conducted one-hour remote interviews via Zoom. To address our RQs, each interview consisted of four major sections.

First, to have a better context building on their pre-screening responses, we asked participants to have a more detailed account of their children's and their VR usage. Then, we asked them about their perceptions of the general risks and benefits of children's VR usage. We did not ask participants about their S\&P-specific perceptions and practices until the next section, because we wanted to see if they would raise any S\&P risks on their own initiative. If not, we were curious about their primary concerns.

Following this section, we explicitly informed participants of our S\&P focus and asked them to specify any S\&P concerns they had. Additionally, we asked participants about their perceived stakeholders and each stakeholder's responsibilities in addressing their concerns.

Last, we asked participants what strategies they had adopted to address the concerns, if at all.  We also asked them about their satisfaction with the existing S\&P controls in VR and their desired features in future VR platforms. We closed the interviews with a live survey about participants' awareness and perceived (or anticipated) effectiveness of the existing parental controls in VR. Before participants filled out the survey, we encouraged them to express their thoughts as they looked at the questions. For all the participants who completed the interviews, we provided one 20 USD Amazon gift card.

\paragraphb{Pilot interviews.} Prior to the formal interviews, we conducted two pilot interviews to assess the interview duration and clarity of interview questions~\cite{majid2017piloting}. The two pilot participants were external researchers with expertise in VR moderation, privacy, and security. One of them was a parent. The pilot participants' relevant research expertise and parental perspectives helped us fine-tune our interview questions. We made sure our interview could be completed within one hour and asked them for advice on our interview protocol. When using the revised protocol in formal interviews, no participants expressed any confusion, indicating the success of pilot interviews. After the pilot interviews, we employed a post-interview survey. However, we found that a live survey using the quiz function in Zoom could be more interactive and useful in eliciting participants' holistic and nuanced thoughts on the features. Therefore, instead of using a post-interview survey, we switched to a live survey as a think-aloud protocol~\cite{jaaskelainen2010think}.


\paragraphb{Data collection and analysis.} We hosted the pre-screening and compensation surveys using Qualtrics~\cite{qualtrics2020}. All the interviews were audio-recorded and transcribed in English using the auto-transcription service provided by Zoom. One researcher carefully checked the accuracy of each transcript prior to the start of the analysis.

Following Saldaña's~\cite{saldana2021coding} coding approach, we conducted an in-depth thematic analysis of our interview transcripts. First, two researchers coded three transcripts independently. After discussing and incorporating the codes, we created an initial codebook that the two researchers agreed on. Using the initial codebook, one of the researchers coded ten of the remaining transcripts, while the other coded the other seven. Similar to many other qualitative S\&P studies~\cite{wermke2022committed, grober2023cloud, holtervennhoff2023wouldn}, the two researchers discussed the codes and resolved any conflicts through several weekly meetings. During our analysis, as we prioritized our primary goal to generate emerging themes and concepts, we did not calculate the inter-rater reliability (IRR) to seek theoretical agreement~\cite{mcdonald2019reliability}.

We created codes about parents' concerns for children's VR activity (e.g., privacy-related, physical health), parents' mitigation strategies (e.g., proactive verbal education), and parents' perceived responsible stakeholders (e.g., VR companies, parents). Then, we merged similar codes and categorized them into emerging themes and sub-themes. Eventually, our codebook consisted of 13 themes, 95 sub-themes, and 613 codes. All the study artifacts, including the final codebook, can be retrieved from Appendix~\ref{sec:appendix_artifacts}.

\paragraphb{Ethical considerations.} This study was approved by the Institutional Review Boards (IRB) of all participating institutions. No personally identifying information was collected from participants. All audio recordings were transcribed and further de-identified immediately after the interviews. We also confirmed participants' consent to participation prior to the interviews and reserved their rights to withdraw at any time. During all stages of data collection, including the pre-screening surveys, interviews, and compensation surveys, participants had the right to withhold any information.

\paragraphb{Limitations.} Given our focus on parents' role in children's S\&P in VR, we only interviewed parents. We must admit that parents' opinions could be skewed without children's input. However, prior work with a similar scope on parents' technology mediation also adopted this approach out of ethical considerations~\cite{cranor2014parents, deldari2023investigation}. Interviewing children and parents from the same family might lead to problems such as intruding on children's privacy from parents, embarrassment and conflicts between children and parents, etc.~\cite{deldari2023investigation}. Considering the ethical conflicts and our focus, we chose to interview only parents and call for future work to investigate children's S\&P perceptions and practices in VR.

As is typical with qualitative studies, our recruited sample size was relatively small. The study relied on participants' self-reported information, which may be subject to social desirability bias. In addition, we did not calculate our IRR. As a consequence, our findings may not be generalizable to all parents of children who use VR. To mitigate this issue, our recruitment was on a rolling basis as mentioned earlier. We summarized the key insights after each interview. Based on the guidelines from prior work~\cite{saunders2018saturation, francis2010adequate}, we identified data saturation after the sixteenth interview, i.e., no more new insights emerged. We then stopped inviting more candidates other than the ones that had already been scheduled. As a result, we conducted four more interviews after identifying saturation, in line with recommended practices~\cite{francis2010adequate}. In doing so, we attempted to cover as many diversified perspectives as possible for an interview study.

%% file: sections/04-results.tex
\section{Findings}
\label{sec:results}


\paragraphb{Parents' demographics.} \label{sec:demographics_parents} Among the 20 participants (see Table~\ref{tab:basic_demographics}), six participants were female and 14 participants were male. The age ranges of participants included 25-34 years old (2), 35-44 years old (3), 45-54 years old (13), and 55-64 years old (2). Participants self-identified as White (11), Black or African American (4), Asian (4), and one participant preferred not to say. In general, parents were highly educated with a bachelor's (6), professional degree beyond bachelor’s (1), master's (9), or doctorate degree (1). One participant had an associate's degree and two participants had no college-level degrees.

Although we did not require participants to specify their occupations, many of them disclosed the industries where they worked at the time of the interview, especially those working with VR/AR technologies and valuing S\&P of technologies. Among the 15 participants who disclosed their professions, seven participants worked in technology companies, many of which focused on VR/AR technology and/or valued the S\&P of their products. Two participants worked in finance and also had some expertise in S\&P. Three participants were K12 educators who had used VR as teaching materials or worked in education technology companies. One participant was a game streamer with abundant experience in VR. The varied experiences of our participants provided us with rich and diversified insights into how parents from different industries and backgrounds approached their children's S\&P in VR. The VR usage details of our study participants and their children are included in Table~\ref{tab:parents_child_usage} in Appendix~\ref{sec:appendix_tables}. The full demographic details are specified in Table~\ref{tab:full_demographics} in Appendix~\ref{sec:appendix_tables}.

\input{tables/participant_demographics}

\paragraphb{In-home VR ownership and parents' usage.} \label{sec:parents_usage} We only interviewed parents who had VR devices in the household. We asked participants to self-describe the VR devices they own in the pre-screening survey and verified the ownership at the beginning of the interview (see Table~\ref{tab:parents_child_usage} in Appendix~\ref{sec:appendix_tables}). Almost all participants (17) only owned one VR headset shared with other family members, including children.

In the pre-screening survey, parents selected their purposes of usage as gaming (16), creativity (8), socialization (8), education (8), streaming and films (7), health (6), and productivity/work (4). During the interviews, we also asked participants how frequently they used VR. Seven participants reported their frequency to be on a daily basis, eight participants were on a weekly basis, three participants were on a quarter basis, and two participants were not actively using VR at the time of the interview but indicated their children were more active users.


\paragraphb{Children's demographics.} \label{sec:demographics_children} Considering children's privacy and the focus of this study, we only asked parents to report the basic demographics of their children and information relating to their S\&P concerns in VR. As shown in Table~\ref{tab:basic_demographics}, 10 children were female and 15 children were male. The age ranges of children included 6-9 years old (6), 10-13 years old (11), and 14-17 years old (8). Therefore, while most VR products are designed for users aged above 13~\cite{Verizon_2023, Gent_LiveScience_2016, Jay_2023}, most children in our sample (17) were below the recommended minimum age for VR usage.

As children of different ages may face different risks in VR, we paid attention to how the age of children may impact parents' perceptions and practices that are closely tied to children's S\&P in VR. Therefore, in our findings, we frequently compare parents of children above and below the recommended age for usage, focusing on their different S\&P concerns and strategies, if any.

\paragraphb{Children's VR access and usage.} \label{sec:children_usage} Regarding children's VR usage, we first asked participants if they gave their children access to VR headsets. All participants said their children had used VR headsets, with varying frequencies. 10 children used VR on a weekly basis, 13 children used VR on a monthly basis, and two children were no longer interested due to the weight of the headset or motion sickness. While most parents (18) said they shared VR headsets with their children, two parents with children aged 14-17 years old said their children had their own headsets.

As shown in Table~\ref{tab:parents_child_usage} in Appendix~\ref{sec:appendix_tables}, according to the parents, their children's purposes of VR usage included gaming (15), creativity (3), education (4), socialization (2), health (2), streaming and watching movies (1), others or unspecified (5).

\paragraphb{Parents' and children's socialization in VR} \label{sec:socialization_vr}
Particularly, since many S\&P risks in VR currently take place during social interaction~\cite{maloney2020anonymity, zheng2023understanding, sykownik2022something}, we also asked participants to specify their socialization experience in VR. 16 participants had at least tried social VR applications or connected with others in VR. Ten participants only met people they already knew in real life, including friends and family members. Six participants socialized with strangers in VR. Notably, one participant was a VR streamer who regularly met many new people in social VR.

By contrast, children's access to social VR was greatly restricted by parents, particularly for younger children. Only three participants whose children were aged 16-17 or 8-9 years old said their children used VR for socialization. One participant whose child was aged 12-13 years old planned to introduce his child to social VR soon. Although most participants said their children had not used VR for socialization, they had observed children's conversations with strangers in VR, mostly during gaming sessions. For children who had access to social or multi-user applications in VR, their parents believed they predominantly or exclusively interacted with children of similar ages or people in their offline connections, e.g., family members, friends, etc.

These details about parents' and children's VR usage are crucial in understanding parents' awareness, attitudes, and coping strategies toward children's S\&P risks in VR, which will be detailed in the remainder of this paper.

\subsection{Parents' Perceptions of Children's S\&P Risks in VR (RQ1)} \label{sec:parents_perceptions}

Due to our interview study's small sample size in qualitative nature and our goal to highlight emerging insights as opposed to generalizability, we adopt a reporting method from a similar interview-based S\&P study ~\cite{emami2019exploring} and use their frequency terminology throughout the remainder of the paper. Instead of using exact numbers of participants, we use ``none'' (0\%), ``a few'' (1\%-24\%), ``some'' (25\%-44\%), ``about half'' (45\%-54\%), ``most'' (55\%-74\%), ``almost all'' (75\%-99\%), and ``all'' (100\%), when appropriate.

In response to RQ1, our findings highlight: (1) Parents generally lack S\&P awareness in VR, largely due to the belief that VR is still in nascent development and therefore risks are a ``future thing.'' (2) S\&P risks are regarded as \textit{potential risks}, whereas the influence of VR on children's physical health is defined as \textit{real-life} risks by parents. (3) Among the identified S\&P risks, parents generally have greater concerns in social interaction contexts than data surveillance by companies. Although they can name a few risks from VR companies, they generally discard such concerns in practice.

\subsubsection{Non-S\&P Concerns as Parents' Primary or Only Concern for Children's VR Usage} \label{sec:non_sp} We asked parents about their concerns toward children's VR usage without priming them with our S\&P focus. As a result, almost no parents brought up concerns for S\&P on their own initiative. Instead, they primarily mentioned concerns for children's physical health issues, online safety, and ``bad influences.''

\paragraphb{Defining eye-sight and physical injuries as real-life risks.} When answering the question about general concerns for VR, most participants illustrated both \textit{risk-life risks} they had perceived or experienced from their own children and \textit{potential risks} they could imagine for other children or in future settings. For real-life risks, most participants mentioned concerns about the influence of VR on children's eyesight. As P15 mentioned, ``\textit{The only risk so far that we've had is my son needs glasses now, but we don't know if there's a direct correlation between the Oculus and his eyesight.}''

Similarly, another participant P2 directly pointed out his ``real-life'' concerns as opposed to ``potential`` concerns:
\begin{myquote}
    ``There's the real-life risk of running into stuff or breaking things as kids get carried away within a particular environment. I don't know if there are long-term risks. There are potential risks like data collection. Ideally, people are collecting data [in VR], but I just don't know if they are.'' 
\end{myquote}

According to P15 and P2, physical injuries were defined as parents' real-life concerns. However, parents also recognized some ``potential risks'' like data collection that they just had no way to directly confirm.

\paragraphb{Stranger danger in a child-adult co-existing ecosystem.} Similar to what prior work has pointed out~\cite{deldari2023investigation, maloney2020complicated, maloney2021stay}, child-adult co-existence in VR was a major safety concern. P1 mentioned:
\begin{myquote}
    ``The gameplay is high-action and fun. So it really appeals to all age groups. So the kids are in there playing with both adults and other kids...I don't think that the companies are doing a good job of limiting it to just kids. They have a financial reason for wanting everybody to use their software.''
\end{myquote}

Some participants echoed P1, indicating their doubt about VR companies' motive and capability to separate children from adult users. According to participants, the child-adult co-existence in the VR ecosystem could lead to many other potential problems, including predators and cyberbullying.

P6, an active social VR user, said, ``\textit{I suppose there could be a way you could track where they (children) were actually if the person was a potential predator. I'm sure there's a way to track the IP of a VR device.}'' P6 had heard of predators collecting IP addresses in social VR, and like some other participants also mentioned, they would be afraid of children being lured to meet online strangers in person, either voluntarily or involuntarily.

Besides predators, participants also mentioned that in an anonymous and immersive environment, cyberbullying could be common and more realistic. As P2 noted, ``\textit{My largest concern would be, the potential for bullying within open environments where you can encounter cyberbullying, particularly with the potential lack of supervision.}'' Like P2, some participants were well aware of the anonymity afforded by avatar-mediated communication~\cite{baker2021avatar, maloney2020talking, fu2023mirror}. They perceived anonymous environments like VR to be a place where cyberbullying was more likely. Worse yet, as P11 stated, ``\textit{[c]yberbullying becomes more visceral in VR (compared to text-based cyberbullying).}''.

\paragraphb{Vigilance toward ``bad influences.''} Besides safety concerns, another frequently mentioned concern was some bad influences from other VR users and the environment designed chiefly for adults.

The first major type of bad influence was bad language in VR observed by participants. A few participants said they had witnessed their children and other children use a lot of bad language, e.g., curse words, racial terms, etc. For instance, participant P16 stated:
\begin{myquote}
    ``I don't like a lot of foul language when they're playing. So that's the one thing I've monitored. If they're playing with people who are using foul language, I tell them to get off." 
\end{myquote}


Besides the under-regulated bad language, the age-mixed environment in VR also brought in the problem of inappropriate content, which has been documented in existing literature~\cite{maloney2020complicated}. Participants, especially those with social VR experience, said that the user-generated worlds were a big issue. Similar to the perception that VR companies did a bad job limiting the age group of users, participants also thought the mechanism of allowing users to upload their own worlds could potentially introduce children to content that was not designed for them.

Importantly, however, participants also noted that bad influences were not only from adults but also from other children who used VR without parents' monitoring. P14, who did not have much social VR experience, said she could imagine ``\textit{kids in VR might have families who have a different or opposite thought process from us.}'' Therefore, P14 was very concerned about the ``\textit{bad influences}'' from other kids in VR.

P14's concerns could be testified by prior work that reported children's bullying behavior in VR~\cite{maloney2020complicated}. Toxic behaviors by children were also spotted by other participants with social VR experience. According to P1, ``\textit{For a few times in public worlds, I've seen that it is obviously fairly young children turned loose in this environment and not being monitored.}'' Echoing P1, P19 even blocked some children in VR. ``\textit{I've learned how to kind of block some children, because it was just, you know, so freewheeling that some children feel like they're uninhibited enough to do anything because you can't see them.}''



In summary, parents were concerned about different types of bad influences from a variety of actors in VR, including adults, children, and user-generated environments.


\subsubsection{S\&P Concerns in Social Interaction Outweighed Data Surveillance} \label{sec:sp_concerns} After asking participants about their general concerns, we specified our focus on children's S\&P in VR. About half of the participants said that they had never thought about the S\&P issues in VR before. After being prompted, participants were able to name some S\&P risks given their prior experience with other technologies. However, some participants would emphasize such concerns hardly occurred to them during children's VR usage. In comparison, while some participants working in technology or S\&P-related industries were more concerned about data surveillance from companies than self-disclosure in social interaction, most of the other participants were more concerned about children's self-disclosure in VR. Mapping this finding to the threat model suggested by Garrido et al.~\cite{garrido2023sok}, we suggest that parents with more S\&P experience are more concerned about hardware and server adversaries. Still, most parents are much more concerned about user adversaries.

\paragraphb{Parents showed awareness of headsets' and apps' data collection risks but discarded the concerns in practice.} For data S\&P in VR, most parents mentioned their concerns for biometric data collection and analyses for user profiling and advertising purposes. Intriguingly, many children and family studies on S\&P-related issues with technology have found that families easily neglect commercial data mining and surveillance in their S\&P mental models~\cite{sun2021they, kumar2017no}. Our findings, however, show some progress that families have made in developing awareness of surveillance capitalism, as our participants covered the four sensitive data types according to Dick~\cite{dick2021balancing}, including observable data, observed data, computed data, and associated data in VR (see Section~\ref{sec:rw1}). For example, P20 said:
\begin{myquote}
    ``[E]ye-tracking would be a big concern of mine. From an advertising and psychology perspective. How people think and what they're looking at are correlated. We would only use products that use eye-tracking internally for system software purposes only."
\end{myquote}

For participants like P20, cutting-edge technology such as eye-tracking could collect physiological and behavioral signals, which could be more invasive compared to their smartphone activity. Echoing P20, P18 mentioned, ``\textit{I would imagine there's a lot more data that comes in when somebody's interacting with VR, so I think the biggest privacy risk is how much neural information you are giving up.}'' Besides ``\textit{neural information,}'' participants were also concerned about their voice data being collected by companies, due to the expansion of AI that could be used to fabricate and manipulate voices.

Importantly, even though participants were aware of the potential data collection and inference by VR companies, they generally discarded such concerns in their parenting practices. As P1 highlighted,

\begin{myquote}
    ``I'm more worried about their interactions with strangers than I am worried about advertisers...There are all kinds of people talking about how eye- and face-tracking can allow companies to know more about you...But I don't know what they are going to do with it that we're worried about. I have never found an answer."
\end{myquote}

Although P1 acknowledged the risks of data collection and inference by companies, he believed that those risks were still very abstract and theoretical to laypeople. He added, ``\textit{A lot of people are doing a good job of presenting these pieces are scary but not doing a good job of telling me why they're scary...But you know the risk of someone knowing your information.}'' We also noticed that a lot of what other participants described as VR data S\&P risks testified to P1's opinion. Participants knew that VR users gave out their biometric data but could not describe the consequences. However, they were able to accurately demonstrate example consequences (as detailed below) of children's self-disclosure in VR.





\paragraphb{Children's self-disclosure in immersive environments was much more alarming.} Participants considered self-disclosure a huge risk for two main reasons: (1) It is hard for children to understand the publicity of having conversations in VR. (2) The nature of immersive environments and ``\textit{cute avatars}'' might trick children into giving out information as naturally as talking with friends in person.

Participants with multiplayer or social VR experience reported their experience encountering children voluntarily giving out information in public worlds. P1 said, ``\textit{If you were playing a game with a kid who knows a credit card number, they would probably just tell you the number. That's hypothetical but I've seen many kids volunteer that sensitive information in VR.}'' P5 mentioned:

\begin{myquote}
    ``In my experience, it doesn't take much to know a kid's private information. Sometimes when I go into rooms, I see kids share all the information publicly and voluntarily...Kids don't understand that it's just like standing on the street corner and shouting it."
\end{myquote}

For P5, the publicity of conversations in VR is hard to be perceived as risky by children as personal information disclosure is too common in VR. 

In addition, some participants attributed children's self-disclosure to the immersion and avatars. P20 said, ``\textit{Feeling immersed in VR might make it a more likely scenario where children might disclose something to someone behind an avatar.}'' Compared to data collection and inference by companies, participants could describe self-disclosure risks in VR in a much more vivid and specific way. They also considered such risks much more ``\textit{direct}'', despite acknowledging the subtle and potentially long-term risks of data surveillance in VR.





\subsubsection{Reasons for Lack of Concerns: Being ``Cutting-Edge'' and ``Harmless''} \label{sec:lacked_concerns} Interestingly, most participants said that they were not concerned about their own children's S\&P in VR at the current stage because VR was still a ``\textit{privilege}'' and there were not many users in it to worry about, especially compared to other more mature technologies like social media. Most parents of teenagers (aged 13-17) also expressed their trust in their children's maturity in using technology. As a result, participants generally lacked S\&P concerns in practice, and therefore did not enforce many mitigation practices.

\paragraphb{Parents lacked the motivation to fight a future thing.} Some participants mentioned that VR was still a privilege in its infancy. For this reason, participants considered VR much less risky than some other technologies that are being used by a majority of people. As P20 explained,

\begin{myquote}
    ``I'm way more concerned about social media...There's not a whole lot of critical mass yet in VR, so things will probably change when everybody piles in."
\end{myquote}

Similarly, a large portion of participants mentioned that the risks they identified would not be concerning until some point in the future when VR technology becomes much more advanced and popular than its current state. VR risks are not ``\textit{well-defined}'' at this point, as P1 said:

\begin{myquote}
    ``The privacy problem in VR is more of a big future thing as they're getting more and more information sometime in the future. It's still a problem that's very nebulous and not well-defined. So I have a hard time being motivated to fight it.''
\end{myquote}

P1 showed a state of mind that is inconsistent with his VR parenting practice, demonstrating \textit{privacy paradox}~\cite{barth2017privacy, wisniewski2018privacy}. Privacy paradox refers to the discrepancies between users' expressed privacy concerns and their actual behavior. In our context, even though P1 knew the risks existed, he still had little motivation to mitigate them in practice.

As a result, participants usually focused more on their children's use of other technologies like social media, neglecting the harms of VR as a ``\textit{future technology}.''

\paragraphb{Trust in children's maturity in technology use.} In addition to defining VR risks as a future thing, some participants,  especially those who were parents of teenagers (aged 13-17), said that they were not concerned because their children were tech-savvy and mature in their S\&P decision-making. As P14 said, ``\textit{I think at this point for him. He's kind of on his path to becoming an adult. So I've given him advice, and I'm kind of trusting that he's gonna follow it. He's fairly mature for 16.}'' Similarly, other parents of teenagers echoed P14 and believed their children's familiarity with VR to have surpassed theirs. 

Some participants who were experienced in VR said that they suspected most other parents would have bought VR just to babysit their children without having any knowledge of VR. As P1 said, ``\textit{I don't think parents understand that there's a shift there. It goes from being something that's happening on a screen in front of you to something that is happening to you.}'' P1 assumed that many parents would consider VR the same as many other technologies like phones, TVs, and tablets.

Indeed, when other participants discussed some risks of VR, they would reference more common technologies, like what P14 (a parent of teenagers) argued, ``\textit{the risks are pretty much the same in any type of technology.}''

Some participants not only neglected the nuances between different technologies, but they also regarded VR as a kind of video game that was just the same as being played on a PC. Interestingly, participants were generally much less concerned about their S\&P when in gaming contexts. As P16, a parent of a teenager nearing adulthood, explained why she did not worry about her son's privacy in VR, ``\textit{I am not concerned because they're just in this game mode. There's nothing personal involved.}'' For some participants like P16, gaming sessions were mostly about play and involved very few privacy risks.

In contrast, parents of children at younger ages (below 13) in our study, such as P2, P7, and P15, showed more vigilance toward VR risks. Most of the parents discussed that they changed or planned to change their strategies as children grew older. Consequently, some parents of older children, especially those nearing adulthood, might be more neglectful of VR risks compared to those of younger children. In general, parents of older children demonstrated an even more imbalanced understanding of VR compared to their children.

Such information asymmetry about technology between parents and children has also been widely documented in existing literature~\cite{cranor2014parents, wisniewski2014adolescent}, impeding parents' ability to engage in teens' technology use. 




\subsection{Parents' S\&P Risk Mitigation Strategies (RQ2)}
Following Wisniewski et al.'s TOSS framework~\cite{wisniewski2017parental}, our findings highlight: (1) Parents primarily rely on active mediation, rules, and restrictions in usage. (2) Parents generally consider parental controls in VR important but face significant technical constraints. (3) In mitigating S\&P risks, parents have to navigate complex tradeoffs between autonomy, safety, and S\&P.

\subsubsection{Active Strategies as the Primary Strategy: Proactively Mitigating Risks via Conversations and Rules} \label{sec:active_strageties} We asked participants what steps they had taken to address the concerns they had, and almost all participants mentioned proactively educating children through conversations and setting rules. However, most of them also emphasized that these conversations and rules were not exclusive to their VR usage. Hence, the ineffectiveness of such strategies in certain VR contexts might have been neglected.

\paragraphb{Parents emphasized online S\&P issues to children, but not specific to VR.} For most families we interviewed, participants claimed that they valued online S\&P education for their children. Having enough conversations about online S\&P was also one of the reasons why participants were not concerned. For instance, P14 said:

\begin{myquote}
    ``I put some rules down for my kids, and they've been pretty good about it. We've had discussions like, we don't put any of your personal information out there. We don't share real names, ages, birthdays."
\end{myquote}



Participants like P14 who said they had strict rules and open conversations with their children all claimed that their children followed their guidance well. However, P14 also emphasized that their rules were ``\textit{pretty much the same across any kind of electronic communication that can go outwards.}''

While participants seemed to be generally confident and satisfied with their approach, P7, who had more VR expertise was opposed to this approach and believed that some physical references should be in place when educating children about S\&P in VR:

\begin{myquote}
    ``We've had those conversations in the context of physical interactions. That makes more sense to kids. When they get on VR, in their head, they're in a safe video game space. They're not thinking, oh, that cartoon character could actually be a person who has ulterior motives like a stranger on a street."
\end{myquote}

According to P7, although parents might have had sufficient privacy education in general digital communication, the proximity between VR and real-life interactions makes it more reasonable to set physical references in their VR privacy education, as opposed to sharing the same set of rules across all digital communication in general.



\paragraphb{Managing children's app access and S\&P information flows in VR through shared and associated accounts.} Unlike social media usage where children usually have separate accounts from parents~\cite{yardi2011social}, most participants shared the only set of VR headsets and accounts with their children. While this practice was not out of mitigation considerations, some participants indicated that sharing headsets and accounts helped them better manage children's VR activity, including their access to certain apps. For example, P14 mentioned:

\begin{myquote}
    ``My kids' accounts are linked to my email, so they have to go through us to buy an app. I can also see any kind of activity. Everything comes to me via email. So I still have a little bit of insight."
\end{myquote}

In some cases, participants made sure to be present when their children tried to create accounts in case they provided excessive information for platforms. As P10 noted, ``\textit{I'm concerned about the phone number and other family information going out. That's why I keep my kids away from creating accounts on their own without having us know.}''

Hence, sharing accounts or creating ones under parents' supervision was a proactive strategy for parents to not only manage children's app access but also their information flows in the VR ecosystem.

Notably, this approach was more discussed by parents of children aged below 13. For parents of children aged above 13, almost all of them did not mention managing or sharing accounts with their children. A few of them mentioned that they would only occasionally check their activity (e.g., interactions with other users) without overly intruding on children's privacy. In addition, the only two participants whose children had their own headsets were aged above 13, further indicating that parents of teenagers may have fewer restrictions and monitoring over their VR usage. This shows a gap where parents of children aged above 13 have limited concerns and mediation toward their children's use of VR products that primarily encourage users aged above 13~\cite{Verizon_2023, Gent_LiveScience_2016, Jay_2023}.

\subsubsection{Challenges in Passive Strategies: Technical Issues \& Parents' Frustrations} \label{sec:passive_strategies} According to the TOSS framework~\cite{wisniewski2017parental}, monitoring and parental controls fall under passive mediation. Our study uncovered that parents faced a range of challenges in their passive strategies, including parents' technical frustrations and children's workarounds.

\paragraphb{Maintaining \textit{ambient awareness}~\footnote{Referencing~\cite{kumar2017no}, we use this term to indicate parents' casual monitoring when they are nearby their children during technology use.} as the most common monitoring strategy.} Almost all participants said they would occasionally check in with children when playing VR, making sure to be able to hear the conversations. In addition, for physical safety considerations, most participants would ask their children to play in open areas like the living room. For example, P11 said:
\begin{myquote}
    ``They basically play in the living room and I'd make sure the volume is loud enough to hear some of the stuff that's being said."
\end{myquote}

However, unless VR is cast on a screen, participants would not be able to see what their children do. While paying attention to the sounds is also a more common option for parents who monitor children using 2D media~\cite{kumar2017no}, VR was much less flexible for our participants who would like to visually check out their children. This constraint influenced parents' practices and attitudes toward different parental controls in VR, as we discuss below.

\paragraphb{Parental controls in VR were important but posed technical constraints.} In the last section of our interviews, we asked participants to first indicate if they knew the feature, and then rate their anticipated effectiveness of each parental control that we found in mainstream VR products. 

\begin{figure}
    \centering
    \includegraphics[width=\linewidth]{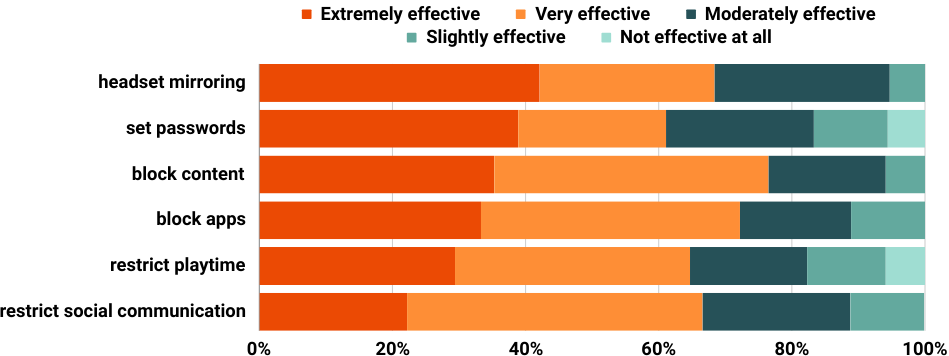}
    \caption{Participants' ratings for the anticipated effectiveness of each parental control feature in VR.}
    \label{fig:controls_rating}
\end{figure}

As a result, 75\% of participants had heard of setting passwords, 70\% had heard of blocking apps, 65\% had heard of headset mirroring and playtime restrictions, 60\% had heard of social communication restrictions, and 55\% had heard of blocking specific content. While this result shows that most participants had heard of these parental controls, their qualitative feedback demonstrates that they had very little experience setting up these controls in practice. For example, P20 said, ``\textit{I don't think I've used parental controls, although I would kind of assume they exist.}''

Some participants like P20 also said that they had never tried setting up parental controls, but given their experience in other technology mediation, they assumed they would be able to set them up.

For the anticipated effectiveness (see Figure~\ref{fig:controls_rating}), overall, headset mirroring was rated as the most effective feature. As mentioned, without casting VR on a screen (i.e., headset mirroring), parents would not be able to visually check out their children's VR activity. Consequently, headset mirroring was rated as the most effective. However, participants who tried headset mirroring highlighted its major drawbacks, including technical issues (e.g., high latency and connection issues) and low engagement for parents. Specifically, two participants said they discontinued the use of headset mirroring because of the unstable connection. For engagement, some participants questioned headset mirroring because they believed it would be unrealistic and not engaging for parents to keep watching children for a long time. As P8 said,

\begin{myquote}
    ``I'm not participating in anything (when looking at the headset mirroring). I'm simply there observing and spectating. So it's of course, not an engaging experience for me. I would rather be doing something where we could do it together. But we don't have a second headset."
\end{myquote}

Besides headset mirroring, blocking apps and blocking content were also highly rated, due to their preventive nature favored by some participants. However, a few participants also pointed out the limitation. As P17 mentioned, ``\textit{Blocking is not going to be effective. There's  always going to be something that you don't know about.}''

To our surprise, while social communication was the most concerning privacy risk to participants, restricting it was not the most effective feature for participants. As P18 explained, ``\textit{I mean, it's a question that, if you want your child to be able to communicate socially, it'll be a kind of an all-or-nothing thing.}'' While restricting social communication could prevent children from interacting with strangers, it was also perceived as too limiting and could not ``\textit{help them learn to do things in the correct way.}'' (P18). Indeed, passive mediation means stopping children from doing certain things and hindering their chances of learning in practice. Navigating such tradeoffs between children's S\&P and their autonomy was a hard process for our participants (as detailed in Section~\ref{sec:autonomy_tradeoffs}). In general, we observed that parents of children aged above 13 were more against the idea of using passive features that could hinder children's autonomy. Hence, some parental controls were much less important to parents of children aged above 13. On the other hand, parents of younger children favored these features for their effectiveness in protecting children from harms.

For passwords and playtime restrictions, while participants found them effective in directly stopping children from accessing VR, they also noted two main drawbacks that discouraged them from using the two controls: (1) By experience, children are likely to work them around and (2) these two controls are too passive while increasing parents' workload, e.g., having to unlock the headset whenever children want to play.


\subsubsection{Complex Tradeoffs Between Autonomy vs. S\&P} \label{sec:autonomy_tradeoffs}
One major discussion we had with our participants was how they weighed and balanced their S\&P mitigation practices and children's autonomy in VR. Participants held two different stances: 1) Some participants would rather trade autonomy for children's S\&P in VR, while 2) most participants advocated for more active approaches to support children's autonomy, even at the cost of potential safety and S\&P risks.

Participants who preferred to trade autonomy for children's S\&P in VR considered the latter more important mainly for two reasons. First, the current VR ecosystem is largely designed for adults, and children are more likely to get into inappropriate situations at this point than later when VR becomes more mature. Hence, it is still too early to give children too much autonomy. Second, VR is still a very minor part of most children's lives, and therefore they have plenty of opportunities to maintain their social autonomy elsewhere, e.g., physical interaction contexts, smartphone usage, etc. As P1 noted:

\begin{myquote}
    ``Today we've gotten to the point where having a phone and a tablet is necessary for you to be on par with your peers...VR is not that way right now...Kids have plenty of other places where they can have autonomy in social experiences...Today, I don't see any problem with VR being completely restricted."
\end{myquote}

On the other hand, participants who advocated for children's autonomy even at the cost of safety and S\&P believed it would be more important for children to gradually learn to tackle harmful situations with parents' guidance. Using passive strategies would hinder children's chances of learning in the real world where parents cannot always keep them company.

Consequently, some participants advocating for the second stance claimed that parents could only help them understand and prepare for potential risks beforehand, whereas when real risks took place, they had to count on themselves with more effective moderation or defensive mechanisms to be in place -- a gap that VR companies have the responsibilities to fill (as detailed in Section~\ref{sec:rq3}). Importantly, most participants also emphasized that the balance between autonomy and children's S\&P and safety in VR would be a dynamic process where the significance of one and the other might be constantly changing in the long run of children's development. Specifically, almost all of the participants who valued S\&P over autonomy had children aged below 13, whereas almost all of the participants whose children were aged above 13 valued autonomy more. This shows the role of children's age in impacting parents' decision-making when facing the autonomy-S\&P tradeoffs.

Overall, parents of children using VR might have to deal with a series of complex and dynamic tradeoffs when trying to mitigate S\&P risks in VR. This finding echoes prior work that suggests managing children's technology is like other daily activities where parents' strategies might be changing in a ``\textit{complex, contextual, and dialectic process}.''~\cite{kumar2017no, mazmanian2017okay}.



\subsection{Responsible Stakeholders and Desired S\&P-Enhancing Features (RQ3)} \label{sec:rq3}
We asked participants to describe the responsibilities of different stakeholders and features that they desired to enhance children's S\&P in VR. Our findings suggest: (1) Participants generally agree a multi-stakeholder ecosystem must be established to make concerted efforts towards more S\&P support for children using VR, with parents being the most accountable stakeholder. (2) While participants desire several features including real-time feedback and alerts with regard to children's S\&P in VR, they also have reservations about VR companies' data surveillance.

\paragraphb{A multisided VR ecosystem towards more S\&P support for children.} When naming perceived stakeholders and their responsibilities in mitigating S\&P risks for children's VR usage, all participants suggested more than one stakeholder. Taking all participants' opinions together, we summarize four main stakeholders -- parents, VR companies (including headsets and apps), governments, and schools/educators.

Among the four parties, parents were identified as the most accountable. Some participants expressed their concerns for adults' lack of S\&P understanding in VR. For instance, P1 mentioned, ``\textit{From what I've seen online, there are a lot of parents handing their kids the VR without even trying VR for a few times. As long as the kid is being quiet and entertained the parents aren't paying attention.}'' To understand the potential risks in VR, participants with more VR experience highly suggested other parents also experience VR, preferably with their children together. For example, P14 demonstrated how she tried VR with her children for the first time:

\begin{myquote}
    ``Let's do it together the first time, so we can see, like what kind of experience and things we need to look out for...Just so we know some potential bad experiences.''
\end{myquote}

P14 said it would be essential for parents to go over the first-time VR experience, gaining some firsthand insights into the potential risks. In doing so, children would be more prepared when harmful situations take place.

Another gap exists in parents' understanding of the S\&P practices and policies across different VR products. We asked our participants if they had ever read any privacy policies on VR platforms, and almost all of them answered no, including those with more relevant expertise. This was in part due to the low readability of the policies, but also because of the inaccessibility of reading them through the headsets.

VR headset manufacturers and app companies (we combine them into ``VR companies'' in this paper) were also considered largely responsible for mitigating the S\&P risks. In our interviews, a few participants expressed their concerns about the conflicts between users' privacy and companies' business models. In general, they believed that most of the existing S\&P-enhancing efforts were primarily for promoting their public image. Correspondingly, prior work has also documented the trade-off that companies face between protecting children's privacy and sustaining their business models~\cite{ekambaranathan2020understanding, ekambaranathan2021money, ekambaranathan2023can}. To mitigate this concern, our participants suggested making S\&P guidance part of a mandatory step before the use. As P18 suggested, ``\textit{I'd prefer if it was in plain language with some short mandatory steps like watching a video... These steps should be very upfront.}'' Making privacy guidance shorter, clearer, mandatory, and upfront would be parents' preferred practice, showing the company's real effort in making the products more S\&P-centered.

Nevertheless, as participants noted, it would be hard for companies to implement such practices as they might conflict with their business models. It would be unrealistic to only count on companies and expect them to be all moral without regulations. Hence, our participants believed that legal regulations must be in place by government bodies to enhance supervision over VR companies' S\&P practices.

Last, as a few participants pointed out, as VR becomes more and more prevalent, schools and educators also have the responsibility in teaching children about the nuances of data surveillance and self-disclosure in VR, compared to other technologies.

\paragraphb{Parents' desired features and S\&P considerations.} The desired features mentioned by participants included (1) real-time S\&P alerts and feedback when harmful interactions take place, (2) AI-based harm detection, (3) a secondary app in place of headset mirroring, and (4) a general usage report sent regularly, e.g., weekly reports, summarizing children's basic VR activity.

As noted by participants, most of the existing moderation and defensive mechanisms in VR are reactive, e.g., blocking and reporting. However, going through the process of blocking and reporting was perceived as ``worthless,'' as it would not make up for the damage and would take too much time to receive feedback. As parents, our participants preferred to know immediately if harmful interactions were taking place. Participants who were concerned about bad influences also hoped to have red flags or alerts sent to them, e.g., an indicator of bad language. However, these ideas also raised some other participants' concerns, mainly for companies' data surveillance. As P18 explained:

\begin{myquote}
    ``While that (real-time alerts) sounds like a really good idea, that would pretty much just be automatically consenting to a hundred percent monitoring of everything that's being said."
\end{myquote}

According to participants, the same concern also applied to AI-based harm detection and secondary apps, all sounding helpful for parents in managing risks more effectively at the price of volunteering more sensitive data to companies.

A few participants proposed an alternative -- sending parents reports with only basic but essential information. In doing so, participants were still able to get an insight into their children's usage while not risking too much of their privacy. As participants did not reach an agreement on how this ``basic but essential'' report mechanism should be designed in the interview, we will further propose our recommendations in Section~\ref{sec:discussion}.

%% file: tables/participant_demographics.tex
\begin{table}
\caption{Parent participants' and their child(ren)'s demographics. (*/: Prefer not to say.)}
\centering
\scriptsize{
\setlength\tabcolsep{8pt}
\begin{tabular}{|l |l |l |l |l |l|} \hline 
\textbf{ID}& \textbf{Age}& \textbf{Gender}& \textbf{Race}& \textbf{Child Ages} & \textbf{Child Genders} \\ \hline 
P1 & 55-64 & M & White & /& F \\  
P2 & 35-44 & M & White & 6-7,8-9 & M,F \\  
P3 & 45-54 & M & Asian & 14-15 & M \\  
P4 & 25-34 & F & Black& 6-7,12-13 & M,F \\  
P5 & 35-44 & M & White & 12-13 & M \\ 
P6 & 45-54 & M & /& 12-13 & /\\  
P7 & 45-54 & M & White & 8-9,10-11 & M \\  
P8 & 55-64 & M & White & 16-17 & M \\ 
P9 & 45-54 & F & White & 10-11,12-13 & M,F \\  
P10 & 45-54 & M & Asian & 10-11,14-15 & M,F \\  
P11 & 45-54 & M & Black& 10-11 & M \\  
P12 & 45-54 & M & Asian & /& M \\ 
P13 & 45-54 & F & White & 12-13,14-15 & F \\ 
P14 & 45-54 & F & Black& 12-13,14-15 & M,F \\ 
P15 & 25-34 & F & White & 8-9 & M,F \\ 
P16 & 45-54 & F & White & 16-17 & M \\  
P17 & 45-54 & M & Asian & 8-9 & F \\ 
P18 & 35-44 & M & White & 16-17 & M \\
P19 & 45-54 & M & Black& 16-17 & F \\ 
P20 & 45-54 & M & White & 12-13 & M \\ \hline

\end{tabular}}

\label{tab:basic_demographics}
\end{table}

%% file: sections/05-discussion.tex
\section{Discussion}
\label{sec:discussion}

In summary, our work has found an alarming status quo where parents of underage VR users generally demonstrate a lack of awareness regarding their children's S\&P in VR due to their conception of VR being ``cutting-edge'' and therefore ``harmless'' (\textbf{RQ1}). Despite parents' lack of S\&P awareness in VR, they carry out a series of general rules toward at-home technology use. Notably, when it comes to VR, parents primarily rely on active strategies as opposed to passive ones that pose significant technical constraints (\textbf{RQ2}). Finally, by prompting our participants in the interviews, they collectively demonstrate a need for a multisided VR S\&P-enhancing ecosystem, with an emphasis on parents' responsibilities and expectations toward a series of S\&P-enhancing features (\textbf{RQ3}). In the remainder of this section, we first discuss how our work extends or contradicts prior work. With the takeaways, we then discuss our recommendations for the critical stakeholders, including parents, educators, VR companies, and governments.

\paragraphb{Parents' knowledge gap in VR led to their overgeneralization of S\&P rules for children in diverse contexts.} While prior work on children in social VR has collectively pointed out that parents have some S\&P concerns for children in VR~\cite{maloney2020complicated, deldari2023investigation, maloney2020virtual}, they did not specify parents' reasons behind the (lacked) concerns and risk mitigation strategies. Interestingly, one 2018 study on VR users' and developers' perceptions of S\&P in VR shows some similar findings, including prioritizing health concerns over privacy and security~\cite{adams2018ethics}. This shows consumers' general S\&P perceptions may not have evolved drastically over the course of the past five years, urging a need to investigate more effective S\&P education, features, and policies regarding VR. Our work builds on this stream of work by first revealing parents' knowledge gap in VR and how it raised S\&P concerns for children, including blurring S\&P rules across a broad range of contexts. 

Prior work has pointed out that parents' incomplete understanding of technology compared to children might impede parents' ability to engage in children's technology use~\cite{cranor2014parents, wisniewski2014adolescent}. As a result, parents might underestimate the amount and type of technology use and online interactions that their children engage in, hence putting them at risk~\cite{ghosh2018safety, blackwell2016managing, wisniewski2017parents}. Similarly, while bystanders in social VR have witnessed great risks for underage users~\cite{deldari2023investigation}, we noticed that parents in our study, especially those with less VR experience, claimed that their perceived S\&P risks would not apply to their own children at the current stage (Section~\ref{sec:non_sp}). Therefore, it is worth suspecting whether the parents had underestimated children's VR activity and S\&P risks. Admittedly, parents' perception of VR's immaturity may be correct in some aspects, especially regarding the immature hardware adversaries~\cite{garrido2023sok}. Most parents without much S\&P experience overlooked or disregarded server adversaries that inferred sensitive data~\cite{garrido2023sok}. In other words, while parents may be correct that immersive data collection had not yet been quite extensive, data inference might have been largely underestimated, according to the established threat model~\cite{garrido2023sok}.

\paragraphb{Parents' overgeneralization of S\&P rules in all digital devices defer addressing concerns to the future.} An important finding of our work is that parents overgeneralize their S\&P rules across diverse contexts, including sharing the same set of rules for all forms of digital communication (Section~\ref{sec:active_strageties}). Despite the fact that parents understand the potential of extensive biometric data being collected in VR compared to other technologies~\cite{sykownik2022something, maloney2020anonymity}, most of them still discard the differences in practice, especially those who seldom or never use internet-enabled VR applications. We believe this phenomenon may be attributed to parents' lack of experience in using VR and their knowledge gap in the nuances between VR and other technologies. We will propose relevant recommendations for parents and educators later in this section.

Another reason for this type of overgeneralization is that parents prioritize children's S\&P in other technologies such as social media (Section~\ref{sec:non_sp}). For this reason, parents deferred addressing their S\&P concerns in VR. Interestingly, in a study of children's S\&P online in general (e.g., social media)~\cite{kumar2017no}, parents still consider S\&P risks as a future concern. Therefore, we suggest that in addition to enhancing parents' understanding of VR technologies, there should be other measures to enhance S\&P education itself more effectively, such as setting physical references as accounted below.

\paragraphb{Parents' perceived differences in S\&P risks between offline and VR settings impede effective S\&P education in VR.} Prior work regarding other consumer technologies has pointed out that parents blur S\&P rules in offline and online settings~\cite{cranor2014parents}. However, despite overgeneralizing S\&P rules across different technologies, most parents in our study understand the shift between S\&P risks in offline and VR settings, mainly due to the anonymity afforded by avatars (Section~\ref{sec:non_sp}). Parents are concerned about children being tricked by the predators behind the attractive avatars.

Problematically, while most parents understand the differences between S\&P risks in offline settings and VR, the proximity in social interaction between VR and face-to-face settings has been overlooked. Only P5 and P7 who have abundant VR experience educate their children about S\&P risks in VR by setting physical references or metaphors (Section~\ref{sec:active_strageties}). All the other parents in our study primarily rely on their general rules for technology use, e.g., not disclosing privacy when chatting with people online. However, text-based communication is drastically different from voice chat in VR~\cite{maloney2020talking}, which resembles talking to friends in face-to-face scenarios. This shows a need to set physical references or metaphors when educating children about S\&P in VR. As suggested by prior work~\cite{li2023s, holvast2007history, malazizi2018risk, regan2002privacy}, physical metaphors are important in improving S\&P awareness and protective strategies. For example, food nutrition labels serve as a physical metaphor for ``privacy nutrition labels,'' better informing users of S\&P-related information~\cite{emami2019exploring, emami2020ask, emami2021privacy}.




\paragraphb{Parents' complex tradeoffs between passive and active strategies raised questions toward S\&P controls in VR.} Our work extends prior work that has discussed children's autonomy and S\&P in their technology use~\cite{ghosh2018safety, wisniewski2017parental, wolfe1978childhood, kumar2017no}, suggesting that parents navigate complex tradeoffs between children's autonomy and S\&P in VR (Section~\ref{sec:autonomy_tradeoffs}). For example, even though parents are largely concerned about children's self-disclosure in social interaction, they consider parental controls that restrict social communication less useful because such an ``\textit{all-or-nothing}'' approach would not help children learn the correct tactics in practice (Section~\ref{sec:passive_strategies}).

Notably, one difference between VR and other consumer-facing technologies lies in the fact that VR has not yet ``\textit{gotten to the point where having a phone and a tablet is necessary for you to be on par with your peers,}'' as highlighted by P1. Due to this perceived difference, some parents consider maintaining children's S\&P much more important than their autonomy in VR. For these parents, passive strategies such as using parental controls are expected to be helpful.

However, even if parents would prefer protecting children's S\&P over autonomy in VR, they are still largely constrained by the existing S\&P controls in VR, due to their low usability, engagement, and granularity (Section~\ref{sec:passive_strategies}). For this issue, we will propose recommendations for VR companies later in this section.

\paragraphb{Recommendation 1: Parents and educators can enhance awareness of technological nuances and physical metaphors in their S\&P education.} As mentioned earlier, parents tend to overgeneralize their S\&P rules in all digital devices (Section~\ref{sec:active_strageties}). Therefore, the first recommendation for both parents and educators (public education for all and educators at school) is to enhance their awareness of the nuances between VR and other technologies that they are familiar with.

For parents, in particular, it is recommended to experience VR themselves as much as possible. It would be even better if parents could maintain their knowledge level about VR on par with their children's. Since VR is an immersive technology, the social experience and data surveillance might be completely different from what they know about smartphones, trying out VR themselves would be the best way for parents to understand the nuances. The significance of parents' firsthand experience in VR is also obvious in our findings, as participants with more VR experience demonstrated better awareness and richer insights.

A specific example is that parents with S\&P experience demonstrated greater concerns toward hardware and server adversaries~\cite{garrido2023sok}. While the lay parents' perception of immature VR technologies may be correct in terms of data collection, their ignorance of the risks of data inference in VR is alarming. Hence, we also urge more privacy education on the different types of adversary types in VR, especially server adversaries.

The same recommendations also apply to educators. As school education has already incorporated S\&P risks in social media as part of the curriculum~\cite{cranor2014parents}, it is also crucial to list VR S\&P risks in the agenda, especially for those who utilize VR as an education technology~\cite{lee2008review, helsel1992virtual, pantelidis1993virtual, Klein_2022}.

When educating children about S\&P risks in VR, both parents and educators should consider leveraging physical metaphors, as opposed to referencing other digital devices. Prior work has suggested that children might perceive physical contexts S\&P riskier than online contexts~\cite{cranor2014parents}. It is therefore highly recommended that parents and educators set more physical references when informing children of the S\&P risks in VR. For example, parents and educators can compare talking in a public social VR world to shouting out loud on a street, as P5 suggested in Section~\ref{sec:sp_concerns}.

\begin{figure}[!t]
    \centering
    \includegraphics[width=0.5\linewidth]{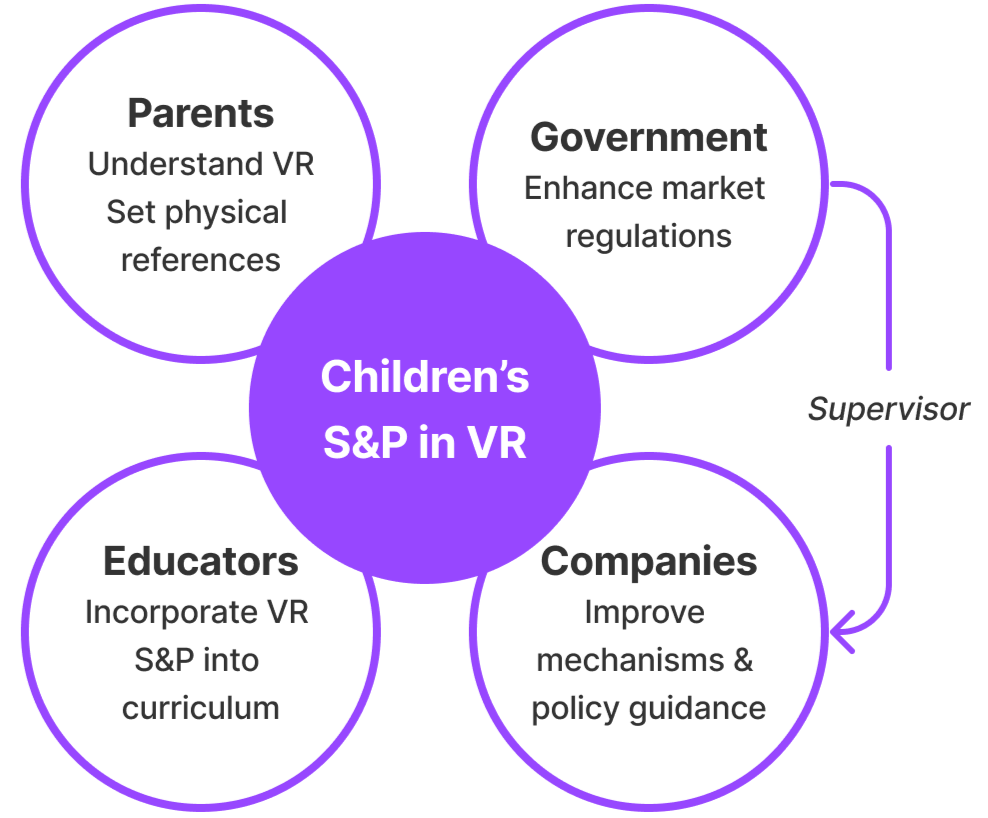}
    \caption{The recommendations for the main stakeholders in the multisided VR ecosystem for children's S\&P.}
    \label{fig:stakeholder}
\end{figure}

\paragraphb{Recommendation 2: VR companies should improve the S\&P controls and guidance for children and parents.} Our findings also highlight that parents generally find S\&P controls unusable for parents and children themselves (Section~\ref{sec:passive_strategies} \&~\ref{sec:rq3}). Parents also find it hard to read the privacy policies in VR (Section~\ref{sec:rq3}). Hence, we call for VR companies to re-design their S\&P controls for both children and parents, involving more \textbf{\textit{engagement, granularity, and modality-specific considerations}}. Drawing on our participants' desired features (Section~\ref{sec:rq3}), we also urge VR companies to provide better S\&P guidance as a part of the mandatory onboarding process.

In improving the usability of S\&P controls in VR, two main problems must be addressed --  (1) technical constraints such as connection issues and high latency, and (2) low engagement. Among a series of controls, parents find headset mirroring most effective (Section~\ref{sec:passive_strategies}). However, due to technical issues, they also refrain from using this feature. In place of headset mirroring, parents hope to use secondary apps accompanied by the VR product (Section~\ref{sec:rq3}). The secondary app can not only replace the functionality of headset mirroring but also provide real-time alerts and other features to help parents gain more engagement while protecting their children in a more timely manner. 

Considering parents' S\&P concerns about these desired features, we suggest increasing the granularity of S\&P controls. Specifically, since parents all have very different considerations when facing the tradeoffs between children's S\&P from the company and strangers, S\&P controls should enable parents to set what exact information they allow companies to collect and analyze for generating real-time or periodic S\&P alerts.

In addition, almost all the parents in our study have never read any privacy policies of VR products, showing an urgent need to improve the S\&P guidance for parents and children in VR. In short, S\&P guidance in VR should be modality-adapted, mandatory, and engaging. Indeed, reading lengthy policies in VR can be difficult~\cite{kojic2020user}, and parents seldom search the S\&P practices of VR products using mobile phones or PCs. Therefore, it is necessary to adjust the design of texts to align with users' reading preferences in VR, considering text lengths, sizes, color contrasts, etc. ~\cite{kojic2020user}.

Besides valuing users' reading experiences in VR, other mechanisms include interactive onboarding tutorials to introduce S\&P-risky scenarios and protective mechanisms. In particular, as prior work suggests~\cite{maloney2020virtual}, a ``\textit{straightforward and well-explained guideline}'' in line with minors' literacy levels is important.

\paragraphb{Recommendation 3: Governments should verify VR companies' S\&P promises without overly hindering immersive experiences.} Even though VR companies demonstrate good S\&P promises through updating their policies, parents still have concerns about whether companies would fulfill the promises (Section~\ref{sec:rq3}). Therefore, it is important for governments or law enforcement sectors to enhance auditing over VR companies' S\&P practices.

On top of enforcing privacy laws to VR~\cite{Reed_2022, Kelly_2022}, prior work has called for policymakers to reform existing privacy laws such as COPPA~\cite{pearlman2021securing} and HIPAA~\cite{gallardo2023speculative} based on the unique affordances of VR such as immersion. While these policies may be generalized to other technologies, they could overly hinder user experiences in VR. Hence, our recommendation is also for the law sectors to enhance their knowledge of VR and iterate privacy laws for VR while supervising VR companies' S\&P practices.

%% file: sections/06-conclusion.tex
This study presents an in-depth qualitative analysis of parents' perceptions and practices toward children's S\&P in VR. Our results highlight that parents generally lack or discard S\&P awareness in children's VR usage. In mediating children's VR usage, parents currently primarily rely on active strategies such as verbal education. Passive strategies, however, have been less effective due to technical constraints. Parents also agree that multiple stakeholders are responsible for mitigating S\&P risks for children, with parents being the most accountable stakeholder. To conclude, we propose specific and actionable recommendations for the critical stakeholders.

%% file: sections/appendix.tex
\newpage
\appendices
\label{sec:appendix}

\onecolumn
\section{Usage and Demographic Tables}
\label{sec:appendix_tables}

\input{tables/parents_vr_usage}

\input{tables/children_vr_usage}
\input{tables/participants_full_demographics}

\section{Study Artifacts}
\label{sec:appendix_artifacts}
{\urlstyle{sf}
\noindent The pre-screening survey, interview protocol, live survey, and final codebook are available at: \\ \url{https://osf.io/4p9c3/?view_only=b1dfae593e5142a6ac0bb59866479d40}.
}

\newpage
\twocolumn[\section{Meta-Review}]

The following meta-review was prepared by the program committee for the 2024 IEEE Symposium on Security and Privacy (S\&P) as part of the review process as detailed in the call for papers.

\subsection{Summary}
This paper investigates parents’ perceptions of security and privacy risks their children face in VR. The authors conduct 20 interviews with parents of children who have used VR to find perceptions of risks, mitigation strategies, and expectations from stakeholders. They find that parents are generally unaware of potential risks, have various protective practices (including active strategies), and request a variety of security \& privacy measures. An extensive discussion of their findings in the context of prior work is included.

\subsection{Scientific Contributions}
\begin{itemize}
\item Provides a Valuable Step Forward in an Established Field
\end{itemize}

\subsection{Reasons for Acceptance}
\begin{enumerate}
\item This paper investigates an important and timely research problem
\item The selection of methods is appropriate and the study is well executed.
\item The authors present interesting findings and an actionable recommendation section.
\end{enumerate}

\subsection{Noteworthy Concerns} 
\begin{enumerate} 
\item Though there is a discussion, the VR threats found in this work aren't strongly tied to VR threats uncovered in prior work.
\end{enumerate}

\section{Response to the Meta-Review} 
When appropriate, we mapped the VR threats in our findings to a previous study as an example~\cite{garrido2023sok} (e.g., Sections~\ref{sec:sp_concerns} \&~\ref{sec:discussion}), calling for raising parents' risk awareness of VR server adversaries. Due to our focus on empirical contributions, we encourage future work to have more in-depth theoretical discussions, mapping the identified VR threats to well-established threat models.


%% file: tables/parents_vr_usage.tex
\begin{table*}[!h]
\caption{Parents'  and Children's VR usage.}
\centering
\scriptsize{
\resizebox{\columnwidth}{!}{%
\begin{tabular}{|l |l |l |l |} \hline 
\textbf{ID} & \textbf{VR Device Ownership} & \textbf{Parent's VR Usage} & \textbf{Children's VR usage.} \\ \hline   
P1 & Meta Quest Pro & Creativity, Education, Gaming, Health, Productivity, Socialization, Streaming & Unspecified \\  
P2 & Meta & Gaming, Work & No longer interested \\
P3 & Meta Quest 2 & Gaming, Exploration of VR & Gaming \\
P4 & Oculus & Creativity, Education, Gaming, Productivity, Socialization & Creativity, Education, Gaming \\
P5 & Meta Quest 2 & Gaming, Health, Socialization, Streaming & Gaming \\
P6 & Meta Quest 2 & Creativity, Gaming, Socialization & No longer interested \\ 
P7 & Meta Quest & Gaming & Gaming \\
P8 & Meta Oculus Quest 2 & Creativity, Education, Gaming, Health, Streaming & Creativity, Education, Gaming, Socialization \\ 
P9 & DEVASCO \& Bnext& Creativity, Education, Health, Virtual Tours & Creativity, Education, Health, Virtual Tours\\
P10 & Oculus, Google Cardboard  & Education & Education, Gaming \\ 
P11 & Oculus, Microsoft Hololens & Creativity, Education, Gaming, Productivity, Work & Gaming \\ 
P12 & Oculus & Creativity, Gaming, Socialization, Streaming & Unspecified \\ 
P13 & Oculus & Gaming, Health & Gaming, Health \\ 
P14 & Oculus Meta Quest & Gaming & Gaming \\ 
P15 & Oculus  & Gaming, Socialization & Gaming, Socialization \\ 
P16 & Oculus Quest 2 & Education, Streaming & Gaming \\ 
P17 & Meta Oculus & Education, Streaming & Gaming, Streaming \\ 
P18 & Oculus  & Gaming & Gaming \\
P19 & Oculus Quest 2 & Education, Gaming, Health, Socialization, Streaming & Gaming \\ 
P20 & Oculus & Creativity, Gaming, Socialization, Streaming & Gaming \\ \hline

\end{tabular}}}
\label{tab:parents_child_usage}
\end{table*}

%% file: tables/participants_full_demographics.tex
\begin{table*}[!h]
\centering
\scriptsize{
\setlength{\tabcolsep}{8pt}
\caption{Parents' age, gender, race, education background, and recruitment channel distributions.}
\label{tab:full_demographics}
\begin{tabular}{|l l |l l |l l |l l|l l|} \hline 
\textbf{Age} &  & \textbf{Gender} &  & \textbf{Race} &  & \textbf{Education Background} & & \textbf{Recruitment Channel} & \\ \hline 
25-34 & 10\% & M & 70\% & White & 55\% & Doctorate degree & 5\% & Personal connections & 5\% \\ 
35-44 & 15\% & F & 30\% & Black & 20\% & Master’s degree & 45\% & LinkedIn & 10\% \\ 
45-54 & 65\% &  &  & Asian & 20\% & Professional degree & 5\% & Facebook & 10\% \\ 
55-64 & 10\% &  &  & Prefer not to say & 5\% & Bachelor’s degree & 30\% & Discord & 10\% \\ 
 &  &  &  &  &  & Associate's degree & 5\% & Reddit & 5\% \\ 
 &  &  &  &  &  & 1 or more years of college credit, no degree & 5\% & School Advertisement & 60\% \\ 
 &  &  &  &  &  & Regular high school diploma & 5\% &  & \\ \hline

\end{tabular}}
\end{table*}